\documentclass[preprint,12pt]{elsarticle}

\usepackage{amssymb,amsmath}
 \usepackage{amsthm}
\usepackage{graphicx}
\usepackage{color}
\usepackage{bbm}
\usepackage{latexsym}
\usepackage{bm}

\newtheorem{defi}{Definition}[section]
\newtheorem{prop}[defi]{Proposition}
\newtheorem{theorem}[defi]{Theorem}
\newtheorem{conj}[defi]{Conjecture}
\newcommand{\beconj}{\begin{conj}}
\newcommand{\enconj}{\end{conj}}
\newcommand{\betheo}{\begin{theorem}}
\newcommand{\entheo}{\end{theorem}}
\newtheorem{condi}[defi]{Condition}
\newcommand{\becondi}{\begin{condi}}
\newcommand{\encondi}{\end{condi}}
\newcommand{\beprop}{\begin{prop}}
\newcommand{\enprop}{\end{prop}}

\newcommand{\bq}{\begin{eqnarray}}
\newcommand{\eq}{\end{eqnarray}}

\def\calH{{\mathcal H }}

\def\C{\mathbb{C}}

\def\I{1_d}
\def\lg{\langle }
\def\rg{\rangle }
\def\deq{\stackrel{\mathrm{def}}{=}}
\def\adg{a^{\dag}}

\newcommand{\ket}[1]{\left\vert\kern.3ex#1\kern.3ex\right\rangle}
\newcommand{\bra}[1]{\langle\kern.3ex #1 \kern.3ex|}
\newcommand{\scalar}[2]{\langle\kern.3ex #1 \kern.3ex|\kern.3ex#2\kern.3ex\rangle}

\def\lu{\mbox{\large 1}}
\def\calP{\mathcal{P}}

\journal{Physics Letters A}

\begin{document}

\begin{frontmatter}



\title{To what extent are canonical and coherent state quantizations physically equivalent?}

 \author{H Bergeron}
 \ead{herve.bergeron@u-psud.fr}
 \address{Universit\'e Paris-Sud, ISMO, UMR 8214 du CNRS, , B\^at. 351, Orsay F-91405, France\fnref{label3}}

\author{J P Gazeau\corref{cor2}}
 \ead{gazeau@apc.univ-paris7.fr}
 \cortext[cor2]{Correspondent author}

\author{A Youssef}
 \ead{ahmed.youssef@apc.univ-paris7.fr}
 \address{APC, Univ Paris Diderot,  Sorbonne Paris Cit\'e,  F-75205, France\fnref{label3}}

\date{\today}                                           




\begin{abstract}
We investigate the consistency of coherent state quantization in regard to physical observations in the non-relativistic (or Galilean) regime. We compare  this particular type of quantization of the complex plane with the  canonical (Weyl) quantization  and  examine whether they are or not equivalent in their predictions.  As far as  only usual  dynamical observables (position, momentum, energy, \dots) are concerned, the quantization through coherent states is proved to be  a perfectly valid alternative. We successfully put to the test the validity of CS quantization in the case of data obtained from vibrational spectroscopy. 

\end{abstract}


\end{frontmatter}








\maketitle

\tableofcontents
\section{Introduction}
\label{intro1}
In this article we examine  fundamental questions about two quantization methods, namely the canonical (or Weyl) and the coherent state quantizations of  non-relativistic systems with one-degree of freedom (to simplify). Are these two procedures equivalent in their predictions?  To what extent can one differentiate between them on a physical level, namely on experimental data, for example those data issued from vibrational spectroscopy \cite{mulliken25,herzberg1989})? \\
The standard (canonical) construction is based on the replacement, in the expression of classical observables $f(q,p)$, of the conjugate variables $(q,p)$ by their respective operator counterparts $(Q,P)$. The latter  obey the canonical commutation rule $[Q,P] = i\hbar \mathbb{I}$. The substitution has to be followed by a symmetrization. The latter step is avoided by the integral version of this method, namely the Weyl or ``phase-space'' quantization (see \cite{phspq06} and references therein).  The canonical procedure is universally accepted due to its numerous experimental validations, one of the most famous and simplest one going back to the early period of Quantum Mechanics with the quantitative prediction of the isotopic effect in vibrational spectra of diatomic molecules  \cite{herzberg1989}. These data validated the canonical quantization, at the opposite of  the  Bohr-Sommerfeld ansatz (no isotopic effect). Nevertheless this does not prove that another method of quantization fails to yield the same prediction. As a matter of fact, we  show in Section  \ref{sec:vibdia} that Coherent State (CS) quantization does provide such a prediction. The latter, also named the  \emph{Berezin-Klauder-Toeplitz}, or \emph{anti-Wick}, rests upon the use of standard (or \emph{Schr\"odinger-Klauder-Glauber-Sudarshan}) coherent states. It consists in taking profit of the resolution, by these states,  of the identity operator to build in a straightforward way quantum observables. CS quantization is very different from the canonical one, and it is not obvious at all that it can eventually yield   predictions compatible with physical evidences. Proving this property is an appealing objective for at least two reasons:
 \begin{itemize}
   \item[(a)] it can validate  CS quantization in general, and allows to implement it  in cases where  canonical quantization is intractable, if not impossible, like for some irregular observables \cite{carnieto68,galapon02} or for distributions  on phase space, in particular when the latter express  geometric constraints, like those encountered in Classical  Gravity.
   
  \item[(b)] since CS quantization produces analytic results of different nature, such as classical-like expectation value formulae (e.g. the so-called lower symbols) for quantum expectation value computations, the former can be considered as exact.
\end{itemize}
The canonical  procedure rests upon the existence of a unique physical constant, namely the Planck constant $\hbar$. It ignores (because of its non-relativistic nature) the natural limitation of Galilean mechanics, exemplified by the existence of a universal speed constant, namely $c$. However, as soon as we deal with the motion of a massive particle, of mass $m$, the existence of $c$ allows to define a natural length in the quantum regime, namely the (reduced) Compton length $\ell_C = \hbar/mc$ \cite{relat}. In fact $\ell_C$ is known as being a (relativistic) localization threshold below which the classical Galilean concept of a particle position loses its validity. Of course this limitation is not usually taken into account  in non-relativistic mechanics. Nonetheless this limitation on localization exists in Physics and induces obvious restrictions on the validity of  non-relativistic mechanics. As is shown in this article CS quantization takes inherently into account this characteristic length within the non-relativistic framework, and so provides a quantum framework which mediates between the non-relativistic domain and the relativistic one (since the speed of light is implicitly involved through $\ell_C$).

  
The article is organized as follows. First we recall the basics of CS quantization, first on a general level (Section \ref{gencsquant}) and next in the particular case where standard CS are employed (Sections \ref{sec:cs} and \ref{sec:csq}). We give in Section \ref{partiCSq} a short insight into the particular place occupied by standard CS quantization among a continuous set of similar integral quantizations including the canonical Weyl transform (\ref{glosspartiCSq} is aimed to give technical details). Physical dimensions are taken into account in Section \ref{sec:csphsp}. The method is applied to a large family of  classical functions in order to obtain explicit formulae (Section \ref{sec:quantpq}). Then we build step by step the CS quantization of a general classical Hamiltonian (Sections \ref{sec:freeH}, \ref{sec:genH}), paying a  special attention to the physical meaning of all parameters and discussing the equivalence of the expressions yielded by CS and canonical quantization.   In Section \ref{sec:quantcomplex} we extend our discussion to the problem of the quantization of general classical functions on phase space. In Section \ref{sec:vibdia} we give a short account of vibrational spectroscopy and related  isotopic effects for diatomic molecules. We use this  example to illustrate  the consistency of the CS quantization with regard to the canonical quantization.
Finally we present in Section \ref{sec:concl} complementary arguments in favor of CS quantization.  

\section{Generalities on coherent state quantization}
\label{gencsquant}
Coherent state quantization is a generic procedure which can be applied to any measure space $(X,\mu)$ (in this sense it could be considered as pertaining to signal analysis too), and \textbf{a fortiori} to any generic phase space. What we need is just a measure and a choice of an orthonormal set. In this sense, it should be distinguished from usual quantization procedures proper to quantum theories, e.g geometric or deformation quantizations whose the mathematical structure, as a departure point,  is  considerably richer. For a clear and comprehensive review of the subject we refer to Ali and Englis in \cite{alienglis2005}.  In CS quantization (see \cite{gazeau2009} for deep probabilistic aspects of the procedure and for a various list of examples, and also \cite{cotgazgor10,bergasiyou10,gasza11,bersiyou12} for recent developments) the approach is simple, of Hilbertian essence, and systematic: one starts from the Hilbert space $L^2(X,\mu)$ of complex square integrable functions on $X$ with respect to the measure $\mu$. One chooses an orthonormal set $\mathcal{O}$ of functions $\phi_n(x)$ in it (set aside the question of evaluation map their respective equivalence classes), satisfying the finiteness and positiveness conditions  $0< \mathcal{N}(x)= \sum_n \vert\phi_n(x)\vert^2 < \infty$ (a.e.), and a ``companion" Hilbert space $\mathcal{H}$ (the space of ``quantum states'')  with orthonormal basis $\{  | e_n\rg \}$ in one-to-one correspondence $\{  | e_n\rg \leftrightarrow \phi_n\}$ with the elements of  $\mathcal{O}$. There results a family $\mathcal{C}$ of unit vectors $|x\rangle$ (the ``coherent states'') in $\mathcal{H}$, which are labelled by elements of $X$ and which resolve the unity operator in $\mathcal{H}$: 
\begin{equation}
\label{defcs}
X\ni x \mapsto | x\rg =\frac{1}{\sqrt{\mathcal{N}(x)}}\sum_{n}\overline{\phi_n(x)} | e_n\rg\, . 
\end{equation}
\begin{equation}
\label{resun}
\lg x| x\rg = 1\, , \quad \int_{X} \mu(dx)\, \mathcal{N}(x)\, | x\rg\lg x| = 1_{\mathcal{H}}\, . 
\end{equation}
This certainly represents the most straightforward way to build ``overcomplete'' or total families of states solving the identity, and one can easily show that most of the various CS families proposed over the last 60 years could have been  obtained in this way.

The resolution of the unity (\ref{resun}) is  the departure point for analysing the original set $X$ and functions living on it from the point of view of the frame (in its true sense) $\mathcal{C}$. More precisely, we understand the map 
\begin{equation*}
f(x) \mapsto A_f \deq \int_{X} \mu(dx)\, \mathcal{N}(x)\, f(x)\, | x\rg\lg x| 
\end{equation*}
as the corresponding CS quantization of $f(x)$. 
We end in general with a non-commutative algebra of operators in $\mathcal{H}$, set aside the usual questions of domains in the infinite dimensional case.  In turn, considering the properties of the new function $\check{f} (x) \deq \lg x| A_f |x\rg$ in comparison with the original $f(x)$ allows to decide if the procedure does or does not make sense  mathematically. 
There is a kind of manifest universality in this approach. Changing the frame family $\mathcal{C}$ produces another quantization, another point of view, possibly mathematically and/or physically equivalent to the previous one, possibly not. 

\section{Standard coherent states with complex parameter\label{sec:cs}}

Let us particularize the above general scheme to the complex plane  and   standard coherent states. The set $X=\C = \left\{z= \dfrac{q+ip}{\sqrt{2}}\right\}$ is the classical phase space for the motion on the line, in appropriate units. It is equipped with the Lebesgue measure $\mu(dx) = \dfrac{d^2 z}{\pi} = \dfrac{dq\,dp}{2\pi}$.
The appropriate choice of an orthonormal set in $L^2(\C,d^2 z/\pi)$ is 
\begin{equation*}
\left\lbrace \phi_n(z) = e^{-\frac{\vert z \vert^2}{2}}\frac{\bar z^n}{\sqrt{n!}}\, , \, n= 0,1, \dotsc\right\rbrace
\end{equation*}
Let $\calH$ be a separable (complex) Hilbert space  with orthonormal basis 
$e_0,e_1,\dotsc, e_n \equiv |e_n\rg, \dotsc$, e.g. the usual Fock space.
To each complex number $z \in \C$ corresponds the unit vector in $\calH$ :
\begin{equation}
\label{stcs}
|z\rg = \sum_{n=0}^{\infty} e^{-\frac{\vert z \vert^2}{2}}\frac{z^n}{\sqrt{n!}}\,  |e_n\rg\, .
\end{equation}
Such a continuous set of vectors is called in the quantum physics literature \emph{family of Glauber-Klauder-Schr\"odinger-Sudarshan coherent states or standard coherent states} \cite{klauskag1985}.
The general construction yield immediately what we expect:
\begin{itemize}
\item[(i)]  Resolution of the unity in $\calH$: 
\begin{equation}
\label{stresunity}
1_{\mathcal{H}}={\displaystyle\int\nolimits_{\C} } \dfrac{d^2 z}{\pi}\,  |z\rg\lg z|\, ,\quad \mbox{in the weak sense}
\end{equation} 
\item[(ii)]  The scalar product or ``overlap'' between two CS,
\begin{equation*}
\lg z| z'\rg = e^{-\frac{1}{2}\vert z - z'\vert^2}\, e^{i\Im(\bar z z')}\, , 
\end{equation*}
 is a reproducing kernel for the Hilbert subspace $\mathcal{K}^L_0$ in $L^2(\C,\, d^2z/(2\pi))$ with orthonormal basis $\bar \phi_n(z) \deq e^{-\vert z \vert^2/2}\,z^n/\sqrt{n!}$: reproducing  (Fock-Bargmann-Segal)  Hilbert subspace. 
 \end{itemize}

%
%
%

\section{CS quantization of functions on the complex plane \label{sec:csq}}

\subsection{Definition}
Thanks to the resolution of the unity or to the underlying reproducing kernel structure, a so-called Berezin-Klauder-Toeplitz-anti-Wick quantization  \cite{alienglis2005,gazeau2009} is made possible, which means that to a function $f(z,\bar{z})$ (or  more generally to a distribution \cite{bechagayo2011}) in the complex plane corresponds the operator $A_f$ in $\calH$ defined by 

\begin{align}
\label{stquant}
\nonumber f \mapsto  A_f &=  \int_{\C}\frac{d^2 z}{\pi}\, f(z,\bar{z}) |z\rg\lg z| \\
 &= \sum_{n, n'= 0}^{\infty}  |e_n\rg\lg  e_{n'}| \,  \frac{1}{\sqrt{n!n'!}}\, \int_{\C}\frac{d^2 z}{\pi}\, f(z,\bar{z}) e^{-\vert z \vert^2}z^n{\bar z}^{n'}\, 
\end{align}
 provided that  weak convergence hold.

\subsection{Symbol interplay}
   
The operator $A_f$ is symmetric if $f(z,\bar{z})$ is real-valued,  bounded if $f(z,\bar{z})$ is bounded, self-adjoint if real semi-bounded (through Friedrich's extension). The original $f(z,\bar{z})$ is an ``upper symbol'' in the sense of Lieb  \cite{lieb1973} or contravariant in the sense of Berezin  \cite{berezin1975}, usually non-unique (except if one restricts the set of functions to be quantized to smooth observables),  for a given operator $A_f$. $f$ is viewed here as a \emph{classical} observable if the so-called ``lower symbol"  $\check{f}$  of $A_f$ 

\begin{eqnarray}
\label{fmapf}
 f \mapsto A_f \mapsto \check{f} (z,\bar z) & \deq \lg z| A_f | z \rg  =  \int_{\C}\frac{d^2 z'}{\pi}\, f(z',\bar{z'}) \vert \lg z| z'\rg \vert^2 \\
\nonumber & = \int_{\C}\frac{d^2 z'}{\pi}\, f(z',\bar{z'}) e^{-\vert z - z'\vert^2}
\end{eqnarray}
has mild functional properties to be made precise (e.g. is a smooth function). The map (\ref{fmapf}) can be viewed as a ``Berezin or heat kernel transform'' \cite{englis1999} in a wide sense.

\subsection{Weyl-Heisenberg algebra}
 
  For the simplest $C^{\infty}$ functions $f(z,\bar{z}) = z$ and $f(z,\bar{z}) = \bar{z}$ we obtain  
\begin{eqnarray}
\label{stoper}
 A_z =  a\, ,  \quad  a\, |e_n\rg = \sqrt{n} |e_{n-1}\rg\, , \quad  a|e_0\rg = 0 \, \quad \mbox{(lowering operator)}\\
 A_{\bar z}  = a^{\dag} \, , \quad a^{\dag} \, |e_n\rg = \sqrt{n +1} |e_{n+1} \rg\quad \mbox{(raising operator)}\, .
\end{eqnarray}  
  These two basic operators obey the so-called canonical commutation rule :
  \begin{equation}
   [a,a^{\dag}] = \I. 
   \end{equation}
   
\section{Comparing Weyl and CS quantizations (and more!)} 
\label{partiCSq} 

 In their seminal paper \cite{cahillglauber69} Cahill and Glauber discuss the problem of expanding an arbitrary operator as an ordered power series in the operators $a$ and $\adg$.  They associate with every complex number $s$ a unique way of ordering all products of these operators. Normal ordering, antinormal ordering (yielded by CS quantization), and symmetric ordering (yielded by Weyl quantization) correspond to the values $s=+1$, $s=-1$, and $s=0$ respectively. Actually, Cahill and Glauber were not directly interested in the question of quantization itself. They start from symmetric operator-valued series $A(a,\adg)$ in terms of powers of $a$ and $\adg$, without considering their classical counterpart from which they could be built from a given quantization procedure, and they ask about their mathematical relations  when a certain $s$-dependent order is chosen.  Nevertheless, their work allows to give a unified view of these different quantizations of the complex plane and of the functions on it.  Let us just summarize here  these interesting aspects in order to mark the specific place occupied  by CS quantization. All details and mathematical justifications are provided by the content of  \ref{glosspartiCSq}.
 
 A general definition including normal, anti-normal and Wigner-Weyl quantizations  for $(X,\nu)= (\C,d^2z/\pi)$ is given by the linear map
\begin{equation}
\label{quantvarpi}
A_f = \int_{\mathbb{C}} \dfrac{d^2 z}{\pi} \varpi(z) \hat{f}(-z) D(z)\, ,
\end{equation}
where $D(z) = \exp(z\adg -\bar z a)$ is the unitary displacement operator, $\varpi(z)$ is a weight function that specifies the type of quantization, and $\hat{f}$ is the symplectic Fourier transform defined by Eq. (\ref{symfourier}).
The map $f \mapsto A_f$ yields the canonical commutation rule $[a,a^{\dag}] = \lu$  if $\varpi$ verifies 
\begin{equation}
\label{regvarpi}
\varpi(0)=1\, , \quad \varpi(-z)=\varpi(z)\, , \quad
\overline{\varpi(z)}=\varpi(z)\,.
\end{equation}
In that case we have 
\begin{equation}
A_{1}=\lu, \, A_{z} = a, \,  A_{\overline{f(z)}} = A_{f(z)}^\dag
\end{equation}
Let us choose with Cahill-Glauber $\varpi_s(z) = e^{s |z|^2/2}$. The parameter $s$ is  the Cahill-Glauber parameter. The case $s=-1$ corresponds to the CS quantization (anti-normal)  since we derive from (\ref{lapD})
\begin{equation}
\label{lapDs-1}
A_f =  \int_{\mathbb{C}} \dfrac{d^2 z}{\pi} f(z) D(z)\ket{e_0} \bra{e_0} D(z)^{\dag}= \int_{\mathbb{C}} \dfrac{d^2 z}{\pi} f(z) \ket{z} \bra{z} \,. 
\end{equation}
The case $s=0$ corresponds to the Weyl quantization since we derive from (\ref{fourD0})
\begin{equation}
\label{fourDs0}
A_f = 2 \int_{\mathbb{C}} \dfrac{d^2 z}{\pi} f(z) D(z) \calP D(z)^{\dag}\,.
\end{equation}
The case $s=1$ is the normal quantization, but its formulation in terms of integral on operators is ill-defined. 
Now, the operator-valued measure (\ref{spovs}) is a positive operator-valued measure iff $s$ real $\leq -1$. This means that the Weyl quantization does not give rise to a real probabilistic distribution on the phase space, as is well known from the definition of the Wigner transform, whereas the CS quantization does it, as we know from the definition of the Huzimi function.

\section{Position and momentum operators from CS quantization\label{sec:csphsp}}
 
\subsection{Coherent states with phase space parameters}
  In order to define harmonic coherent states $\ket{ \xi_{q,p} }$ that live on the physical classical phase space $\mathcal{P}=\{ (q,p) \in \mathbb{R}^2 \}$ we need to introduce {\it two} physical constants (and not a unique one as is the case of canonical quantization):  the reduced Planck constant $\hbar$ and an arbitrary length $\ell$. Equivalently, we can introduce $\ell$ and the momentum parameter $\wp=\hbar/\ell$. Then we define the normalized vectors $\ket{ \xi_{q,p} }$ from the states $\ket{z}$ of  \eqref{stcs} as
\begin{equation}
\ket{\xi_{q,p}}= \ket{\frac{q}{\ell \sqrt{2} } + i \frac{p \ell}{ \hbar \sqrt{2}} } = \ket{\frac{q}{\ell \sqrt{2} } + i \frac{p}{ \wp \sqrt{2}} }\, .
\end{equation}
  The resolution of unity in \eqref{stresunity} becomes
\begin{equation}
\int_{\mathcal{P}} \frac{dq dp}{2 \pi \hbar} \ket{\xi_{q,p}} \bra{\xi_{q,p}} = \mathbb{I}\, .
\end{equation}

\subsection{Position and momentum}
   The CS quantization of the classical observables $q$ and $p$ leads to 
 \begin{equation}
  Q= A_q=\frac{\ell}{\sqrt{2}}(a+a^\dagger) \quad {\rm and} \quad P=A_p=\frac{\hbar}{i \ell \sqrt{2}} (a-a^\dagger),
 \end{equation}
   where $a$ and $a^\dagger$ are the operators previously defined. Operators $P$ and $Q$ verify the canonical rule $[Q,P]=i \hbar \mathbb{I}$ and therefore correspond to the operators of usual canonical quantization.
 
Let us remark at this stage that $\ell$ is a free parameter of the theory without well-defined physical meaning, since on a physical point of view, only the spectra of  the operators $P$ and $Q$ are observables.
 
\section{CS quantization of functions of $q$ or $p$ \label{sec:quantpq}}
In this section, we only present the mathematical expressions and their main features. The physical analysis is developed in the following sections.

\subsection{Functions of $q$}
For a general classical function $f(q)$  the corresponding operator $A_{f(q)}$ is easily proved to be the following convolution:
\begin{equation}
\label{eqn:fq0}
A_{f(q)}=\tilde{f}(Q) \quad {\rm with} \quad \tilde{f}(q)= \int_{\mathbb{R}} f(q-x) e^{-x^2/\ell^2} \frac{{\rm d}x}{\sqrt{\pi \ell^2}}\,.
\end{equation}
This expression stems from  the well-known ``Gaussian $x$-representation" $\scalar{\delta_x}{\xi_{p,q}}$ of coherent states and then from  the ``matrix elements" $\bra{\delta_x} A_{V(q)} \ket{\delta_y}$.

For $\ell$ small enough, we have $e^{-x^2/\ell^2}/\sqrt{\pi \ell^2} \simeq \delta(x)$, and then we recover $\tilde{f}(Q) \simeq f(Q)$. The first correction is given by a second derivative of $f$
\begin{equation}
\label{eqn:fq1}
\tilde{f}(Q) = f(Q) +\frac{\ell^2}{4} f''(Q)+ o(\ell^2)\,.
\end{equation}
We conclude that the operators given by CS and canonical quantization (respectively $\tilde{f}(Q)$ and $f(Q)$) coincide in the limit $\ell \to 0$.

\subsection{Functions of $p$}
For a general classical function $f(p)$ we have similarly:
\begin{equation}
\label{eqn:fp0}
A_{f(p)}=\tilde{f}(P) \quad {\rm with} \quad \tilde{f}(p)= \int_{\mathbb{R}} f(p-x) e^{-x^2/\wp^2} \frac{{\rm d}x}{\sqrt{\pi \wp^2}}\,.
\end{equation}
where $\wp=\hbar/\ell$. For $\wp$ small enough, we have $e^{-x^2/\wp^2}/\sqrt{\pi \wp^2} \simeq \delta(x)$, and then we recover $\tilde{f}(P) \simeq f(P)$. The first correction is given by a second derivative of $f$
\begin{equation}
\label{eqn:fp1}
\tilde{f}(P) = f(P) +\frac{\wp^2}{4} f''(P)+ o(\wp^2)\,.
\end{equation}
We conclude that the operators given by CS and canonical quantization (respectively $\tilde{f}(P)$ and $f(P)$) coincide in the limit $\wp \to 0$.

\subsection{More general cases}
Still using the Gaussian position representation or momentum representation of coherent states, we obtain the following explicit formulae for the CS quantization of classical functions $p f(q)$ and $q f(p)$

\begin{eqnarray}
\label{eqn:mixedfqp}
A_{p f(q)}= \dfrac{1}{2} (P \tilde{f}(Q) +  \tilde{f}(Q) P) \\
A_{q f(p)}= \dfrac{1}{2} (Q \tilde{f}(P) +  \tilde{f}(P) Q)
\end{eqnarray}
where the functions $\tilde{f}(Q)$ and $\tilde{f}(P)$ are  defined by Eqs. (\ref{eqn:fq1}) and Eqs. (\ref{eqn:fp0}) respectively. In particular
\begin{equation}
A_{p q}=\dfrac{1}{2}(P Q + Q P)
\end{equation}

\subsection{Remarks about the limits $\ell \to 0$ and $\wp \to 0$}
First of all, these two limits lie beyond the validity range of CS quantization, since  both the conditions $\ell \ne 0$ and $\wp \ne 0$ are  required for defining the coherent states. 

Secondly, these states, say in position representation,  do not become at the limit $\ell\to 0$ \emph{the unnormalizable coordinate states $\delta_x$ on which canonical quantization is based}, even though the squared modulus $\vert \lg \delta_x| z\rg\vert^2$ converges to the Dirac distribution $\delta(x - q)$ where $q$ is the position parameter of the CS. Actually this limit does not imply that the projector $|z\rg \lg z|$ tends to $\delta (Q-q) \equiv |\delta_q\rg\lg \delta_q|$. In fact the expression $\lg \delta_x| z\rg \lg z| \delta_y \rg \to 0$ in the $\mathbb{R}^2$-distribution sense.  \textit{Mutatis mutandis}  the same  holds true for the limit $\wp \to 0$.

Finally, let us remark that the constraint $\ell \wp=\hbar$ must be preserved as far as we want to remain in the quantum domain. Therefore both limits $\ell \to 0$ and $\wp \to 0$ cannot be carried out at once. Consequently CS and canonical quantizations are not \emph{mathematically} equivalent, i.e. they do not map  in general a classical function to  the same operator  (even in some limit sense).

Nonetheless, this conclusion does not mean that CS and canonical quantizations are not \emph{physically} equivalent (within the specific non-relativistic framework, a premise of the present study), i.e. do not provide the same \emph{measurable} predictions (up to usual experimental uncertainties). As  is shown in the sequel, this apparent paradox can be overcome by using two  arguments: (a) the existence of superselection rules that prevent certain self-adjoint operators from being physical (i.e. \emph{measurable}) observables, (b) the intermediate physical status of non-relativistic mechanics that prohibits non-relativistic expressions to be physically \emph{exact} (as approximations of relativistic ones).

\section{The quantized free Hamiltonian and the parameter $\ell$ \label{sec:freeH}}

\subsection{Classical and quantum free Hamiltonians}
We now introduce the mass $m$ of the particle, the latter being 
viewed on the classical level as a material point. The classical free Hamiltonian is usually defined as $h_0=p^2/2m$. This definition ignores a possible constant additive term, namely the proper energy $\mathcal{E}_0$ of the particle, that is the energy of the particle in its rest frame. This term is usually omitted because it is non-measurable in the classical Galilean  framework. (Of course classical quantities are never really measurable since the real world is quantum, but we claim that $\mathcal{E}_0$ is already not  measurable in the classical context.) This ``classical non-measurability'' is due to : (a) the absolute conservation of $\mathcal{E}_0$ \cite{leblond1974}  for a given particle (in other words $\mathcal{E}_0$ is necessarily a classical particle-dependent physical constant), (b) there is no absolute energy origin in the Galilean framework (therefore only particle energy variations are classically measurable). Then, on a Galilean standpoint, we are free to choose the energy origin as being exactly the energy of the particle in its rest frame, and then $\mathcal{E}_0=0$, but this is only one choice between others.\\
Now, let us assume that we decide to take into account this proper energy term, labelled $\mathcal{E}_0^{(C)}$, in order to specify its classical origin. The classical non-relativistic Hamiltonian becomes $h_0=p^2/2m+\mathcal{E}_0^{(C)}$, while from \eqref{eqn:fp0} its CS-quantized version $H_0=A_{h_0}$ reads as
\begin{equation}
H_0=A_{h_0}=\frac{P^2}{2m}+ \frac{\hbar^2}{4 m \ell^2}+\mathcal{E}_0^{(C)} \, .
\end{equation}
  The quantum proper energy $\mathcal{E}_0^{(Q)}$ of the particle, defined as the infimum of the spectrum of $H_0$, is $\mathcal{E}_0^{(Q)} = \inf {\rm Sp} (H_0)=\dfrac{\hbar^2}{4 m \ell^2}+\mathcal{E}_0^{(C)}$ which differs from  $\inf h_0(p) =\mathcal{E}_0^{(C)}$. Therefore CS quantization generates an additional proper energy$\dfrac{\hbar^2}{4 m \ell^2}$, whereas usual canonical quantization preserves proper energy according to \\
  $\inf {\rm Sp} (P^2/2m +\mathcal{E}_0^{(C)})=\mathcal{E}_0^{(C)} = \inf h_0(p)$. 
  
  First we observe that this result yielded by CS quantization appears as a renormalization of {\it bare} parameters obtained in quantum field theory. Second, through this property we recover some features of the old idea of ``proper wave function" of a particle first introduced by Lande \cite{lande1939} and Born \cite{born1939} (see also \cite{stali85}) to describe elementary particles that are not point-like. Third, this seems to highlight an obvious difference between CS and canonical quantization. But in fact this conclusion holds only on a mathematical level, not on a physical one. 
  
Exactly as {\it bare} parameters are not physically measurable in quantum field theory, the classical proper energy $\mathcal{E}_0^{(C)}$ is not physically relevant: only the {\it quantum} proper energy is \textbf{possibly} meaningful (depending on its measurable nature). Therefore the previous result  means that the non-measurable parameter $\mathcal{E}_0^{(C)}$ must be adjusted differently in CS quantization and in canonical one, in order to provide the same quantum proper energy $\mathcal{E}_0^{(Q)}$. Actually since the absence of absolute energy origin in a Galilean framework remains valid on the quantum level, $\mathcal{E}_0^{(Q)}$ is  not measurable in any fashion, and we are free to remove it. 
  
The conclusion of this analysis is that the mathematical difference between the expressions issued from CS and canonical quantization does not  result in incompatible physical predictions in the case of the free particle.
  
\subsection{The parameter $\ell$}

\subsubsection{First analysis}
Due to the fact that CS quantization needs two physical parameters $\ell$ and $\hbar$, and not uniquely $\hbar$ as it holds in canonical quantization, two approaches are possible. The first one consists in assuming that $\ell$ is a purely mathematical parameter that must be eliminated at the very end of the procedure by taking some limit (e.g. $\ell \to 0$). The second one is to consider that $\ell$ is really a physical parameter. Such a quantity is neglected when the canonical quantization is worked out. Indeed, in non-relativistic mechanics the numerical effect of $\ell$  is negligible on \emph{concretely measurable} quantum observables. In the following we analyze this last option and we examine the possible order of magnitude of $\ell$. But we need first to study the possible dependences on $\ell$. Since the purpose of CS quantization is to provide a linear procedure giving the quantum version of all phase space classical observables corresponding to a given non-relativistic particle, the parameter $\ell$  must be either an absolute length, or a particle-dependent length. Nevertheless,  in any case (due to the linearity assumption) $\ell$ cannot be an observable-dependent parameter (in particular a potential-dependent one). Furthermore since there is no natural absolute length (above the Planck scale), we investigate the particle-dependent possibility after proceeding with some dimensional analysis.

Inasmuch as CS quantization involves the two constants $\ell$ and $\hbar$, as soon as we introduce the mass $m$ of a particle, we can build a new natural constant which is speed dimensional $v=\dfrac{\hbar}{m \ell}$. Since there is a unique fundamental unit of velocity in particle physics, we are led to identify $v$ with the speed of light $c$ (up to some numerical factor). Therefore the order of magnitude of $\ell$ naturally appears as being the Compton length $\ell_C=\dfrac{\hbar}{m c}$.  As a consequence,  CS quantization provides some intermediate quantum framework between the non-relativistic domain and the relativistic one (presence of $c$ in  $\ell_C$), while usual canonical quantization is strictly limited to the non-relativistic sector. Of course this conjecture must be supported by more investigations.  The latter are detailed in the sequel.


\subsubsection{The parameter $\ell$ and the Compton length}
The reasoning developed above for the free Hamiltonian  provides another argument in favor of  the Compton length scale. Let us decide to retain the quantum proper energy term $\mathcal{E}_0^{(Q)}=\frac{\hbar^2}{4 m \ell^2}+\mathcal{E}_0^{(C)}$, choosing now the relativistic energy origin. Then $\mathcal{E}_0^{(Q)}=\frac{\hbar^2}{4 m \ell^2}+\mathcal{E}_0^{(C)}= mc^2$. Taking $\mathcal{E}_0^{(C)}=0$, we deduce $\ell=\ell_C/2$. Actually this reasoning does not prove  that the free parameter $\ell$ {\it is} exactly one-half of the Compton length, but it proves that this choice of $\ell$ (and of $\mathcal{E}_0^{(C)}$) is compatible with the relativistic proper energy (which is of course beyond the range of the non-relativistic domain). 
  

  A last argument in favor of the choice of $\ell_C/2$ as the natural order of magnitude of  $\ell$ is based on the Heisenberg inequality:  if we assume that the particle is localized in an interval of length $\Delta x$ smaller than $\ell_C/2$,  the uncertainty on velocity $\Delta v$ becomes greater than $c$, which is physically irrelevant. In fact $\ell_C$ is known as being a (relativistic) localization threshold below which the classical Galilean concept of a particle position loses its validity \cite{relat}. Although this limitation is not usually taken into account in the non-relativistic context, this limitation does exist and it induces precise limitations to non-relativistic mechanics. For example a classical potential energy $V(q)$ considered in the non-relativistic domain and varying  significantly on the scale of $\ell_C$ is physically irrelevant (such a potential belongs to the relativistic framework). 
  
 \subsubsection{Conclusion to this section}
  From the previous analysis, we find reasonable to assume that the length $\ell$ is a particle-dependent parameter with the same order of magnitude as the Compton length. The following section devoted to the quantization of potential energy  reinforces this assumption. Therefore we fix from now on $\ell$ as being $\ell=\ell_C/2$ (and $\mathcal{E}_0^{(C)}=0$) to obtain explicit formulae compatible with the relativistic proper energy (even if only  the magnitude order is  important). Consequently in the sequel the free Hamiltonian $H_0$ is written as
  \begin{equation}
  \label{eqn:freeH}
  H_0 = A_{p^2/2m} = \dfrac{P^2}{2m} + m c^2\,  .
  \end{equation}

\section{The quantized Hamiltonian: 1D general case \label{sec:genH}}

\subsection{Classical and quantized potential energies}
We know that  $V(q)$ is defined up to a constant. Therefore a supplementary condition must be added to fix unambiguously the function $V(q)$ (e.g. its  value  at its minimum, or at infinity). By using \eqref{eqn:fq0} we obtain its quantum counterpart

\begin{equation}
\label{eqn:quantv}
A_{V(q)}=\tilde{V}(Q) \quad {\rm with} \quad \tilde{V}(q)= \int_{\mathbb{R}} V(q-x) e^{-x^2/\ell^2} \frac{{\rm d}x}{\sqrt{\pi \ell^2}}\,.
\end{equation}
For $\ell$ small enough, we can also use \eqref{eqn:fq1}, to obtain the approximation
\begin{equation}
\tilde{V}(Q) = V(Q) +\frac{\ell^2}{4} V''(Q)+ o(\ell^2)\,.
\end{equation}
As was previously mentioned a classical potential energy in the non-relativistic domain cannot possess significant variations at the $\ell_C$-scale. This requirement justifies the above expansion. Replacing $\ell$ by $\ell_C/2$ we obtain

\begin{equation}
\label{eqn:qpot}
\tilde{V}(Q) \simeq V(Q) +\dfrac{\hbar^2}{16 m^2 c^2} V''(Q)
\end{equation}
Comparing this expression with the one obtained from canonical quantization, we see that CS quantization yields the additional  $\frac{\ell^2}{4} V''(Q)$. This term is a correction always {\it physically} negligible in the non-relativistic domain. Indeed it {\it looks like} a relativistic correction, namely the so-called Darwin term resulting from the Foldy-Wouthuysen transformation \cite{itzykson1980}. 

Once more we observe that the formulae obtained from CS and canonical quantization respectively  differ on a mathematical level. Neverheless the terms involved in this difference cannot lead to physical inconsistencies in the non-relativistic domain.

\emph{Remark} The expansion \eqref{eqn:qpot} is valid if $V$ is smooth enough. In the case of discontinuous potentials, even in the case of some singular potentials ($V(q)=1/\sqrt{|q|}$), the interest of the quantization by convolution \eqref{eqn:quantv} is that its application range is considerably wider than the canonical one.

\subsection{Hamiltonian with a magnetic field}
In the case of a ``magnetic'' potential introduced via the usual minimal coupling the classical Hamiltonian reads $h(q,p)=\dfrac{1}{2m} (p-e A(q))^2$. By using  \eqref{eqn:fq0}, \eqref{eqn:fq1}, \eqref{eqn:mixedfqp}, \eqref{eqn:freeH}, we obtain the CS-quantized version $H$  as
\begin{equation}
H = \dfrac{1}{2m} (P-e \tilde{A}(Q) )^2 + \dfrac{e^2}{2m} (\widetilde{A^2}(Q)-\tilde{A}(Q)^2) + mc^2
\end{equation}
and then

\begin{equation}
H \simeq \dfrac{1}{2m} \left( P-e A(Q)- e \dfrac{\ell^2}{4} A''(Q) \right)^2 + \dfrac{e^2}{4m} \ell^2 A'(Q)^2 + mc^2 + o(\ell^2)
\end{equation}
with $\ell=\ell_C/2$. We observe once more that the additional terms issued from CS quantization induce negligible effects in the non-relativistic regime.

 Note that in one dimension all magnetic potentials $A(q)$ are gauge fields and the present discussion has a physical interest  in its 3D generalization only.
 
\subsection{The general case}
Collecting the results of the previous sections,  the general quantum Hamiltonian $H$ generated by CS quantization and associated to the classical potentials $A(q)$ and $V(q)$ reads as
\begin{equation}
H = \dfrac{1}{2m} (P-e \tilde{A}(Q) )^2 + \tilde{V}(Q) + \dfrac{e^2}{2m} (\widetilde{A^2}(Q)-\tilde{A}(Q)^2) + mc^2 \,.
\end{equation}

\subsection{The universal  example: quantum harmonic oscillator \label{sec:csho}}
 The classical harmonic potential is $v(q)= \frac{1}{2} k q^2 \equiv \frac{1}{2} m \omega^2 q^2$ where $k$ is the constant force and $\omega$ is the usual vibrational parameter.  The CS quantized counterpart $V=A_{v(q)}$ of $v(q)$ reads exactly as
\begin{equation}
V=A_{v(q)}= \frac{1}{2} m \omega^2 Q^2 +\frac{1}{4} m \omega^2 \ell^2\,.
\end{equation}
   Hence, the CS quantization of the classical Hamiltonian $h(p,q)=p^2/2m+v(q)$ leads to the quantum Hamiltonian $H$

\begin{equation}
H=  \frac{P^2}{2m} + \frac{1}{2} m \omega^2 Q^2 + mc^2+ \gamma\hbar \omega\, . 
\end{equation}
Here  $\gamma = \frac{\hbar \omega}{16 mc^2}$ is a dimensionless factor expressing the ratio between two typical energies, namely the (non-relativistic) quantum energy $\hbar \omega$ and the rest mass of the particle. The eigenvalues $E_n$ of $H$ are
\begin{equation}
E_n = (n+\dfrac{1}{2}+\gamma) \hbar \omega + mc^2
\end{equation}
Since the validity of the classical Hamiltonian $h(p,q)$ is restricted to the non-relativistic domain, the ratio $\gamma$ is extremely small and can be neglected in that context. Therefore the Hamiltonian reduces to the one yielded by canonical quantization, up to the relativistic proper energy of the particle.

\section{Quantization of more elaborate ``observables'' \label{sec:quantcomplex}}
The \emph{physical} relevance of the quantization of general classical ``observables'' $f(p,q)$ (through either the CS or the canonical method) is a question that must be carefully analyzed. On a mathematical level, it is obvious that these two procedures provide generically different operators for the same classical function (the previous sections already prove these differences). But as was previously observed for a class of concrete dynamical observables (position, momentum, energy, \dots), the order of magnitude of these mathematical differences is not sufficient to produce significant physical differences. Let us call $\mathcal{A}_P$ this subset of quantum operators. $\mathcal{A}_P$ is only a subset of the set of all possible self-adjoint operators. So a fundamental question arises: is any self-adjoint operator  a ``physical observable''? In other words, looking at the non-relativistic behavior of a particle,  say on the quantum level, and given any mathematical ``observable'', that is a self-adjoint operator $A$, are we {\bf really} able to propose a {\bf concrete} experimental setup adequate to measure the spectrum of $A$ (even if we assume the existence of devices which are not realizable today)? If we answer positively  this question, then CS and canonical quantization are certainly not \emph{physically} equivalent. If we answer negatively, CS and canonical quantization could be \emph{physically} equivalent if the \emph{measurable} observables precisely belong to the subset $\mathcal{A}_P$. This assumption corresponds to a kind of superselection rule within the non-relativistic framework.



\section{The observed isotopic effect in vibrational spectra of diatomic molecules}
\label{sec:vibdia}

As was previously shown, a superficial analysis of CS quantization might lead to wrong conclusions, due to a poor consideration of the role played by the choice of a zero for the energies within a non-relativistic physical framework. Indeed an absolute value of energy cannot be defined in  (classical or quantum) non-relativistic mechanics, since the ``proper energy" of a particle is not measurable. Then the fact that the energy spectrum predicted by CS quantization always differs at least by a constant from the spectrum yielded by canonical quantization (see section \ref{sec:csho} for the harmonic oscillator) is not sufficient to conclude that CS quantization must be rejected: this just means that CS quantization generates a ``quantum proper energy" for the particle, while canonical quantization does not. In fact, as we prove below, CS and canonical quantizations predict the same correct isotopic effect.

\subsection{The physical context}

  The observed spectrum of diatomic molecules is mainly composed of vibration  (near infrared) and rotation (far infrared) spectra. The condition for the occurrence of 
  the former one  is that the rate $M_1$ of change of the dipole moment $M \approx M_0 + M_1 x$ is different of 0. This condition is fulfilled only for molecules with unequal nuclei.
 
     When dealing with the electronic transitions of diatomic molecules one considers $E_e \equiv E^{\mathrm{el}} + V_{\mathrm{nuc}}$ where 
  \begin{itemize}
   \item $E^{\mathrm{el}}= E^{\mathrm{el}}(R)$ is an eigenvalue of the electronic part  of the Schr\"odinger equation at fixed nuclei interdistance $R$, within the framework of the Born-Oppenheimer approximation, 
   \item and $V_{\mathrm{nuc}} = V_{\mathrm{nuc}}(R)$ is the Coulomb potential of the nuclei. 
 \end{itemize}  
    
   Therefore one considers $E_e(R)\equiv U(q)$, with $q\equiv R -R_{\mathrm{eq}}$ (where $R_{\mathrm{eq}}$ is the equilibrium interdistance corresponding to the minimum of $E_e(R)$) as the vibrational potential energy of the nuclei issued from a given  stable  electronic  state. 
    
     The minimum of the lowest (i.e. ground) electronic state, as shown in Figure \ref{figure1}, \textbf{is usually chosen as the origin of the energy range}.
    
      Each electronic potential curve gives rise to a quantum vibrational Hamiltonian (obtained from  canonical quantization) that leads to a spectrum of eigenvalues. The latter are the harmonic ones ($\hbar \omega_e (n+ 1/2)$) + successive anharmonic corrections ($-\hbar \omega_e x_e (n+ 1/2)^2 + \hbar \omega_e y_e (n+ 1/2)^3 + \cdots$), $\omega_e  \gg \omega_e x_e  \gg \omega_e y_e$,  with the  notations used in \cite{herzberg1989}). Since different electronic states correspond to different potential curves, the parameters $\omega_e$, $x_e$, $y_e$ are different in each case.
     
     The observed vibrational spectra correspond either to transitions between two vibrational levels corresponding to a single electronic state (typically the ground electronic state), or to transitions between vibrational levels corresponding to two electronic states (the ground electronic state and an excited one). The energetic diagram is represented in Figure  \ref{figure1}.

  \subsection{The predicted spectra from canonical quantization}
 
  \subsubsection{Notations (with $\hbar = 1$) and actual measurements concerning a single electronic state (the ground electronic state)}
   
  The vibrational energies $E_n$ are expressed as 
\begin{equation}
 E_n=G \left(n+\frac{1}{2} \right)=\omega_e (n+ 1/2)- \omega_e x_e (n+ 1/2)^2 + \omega_e y_e (n+ 1/2)^3 + \cdots.
\end{equation}
   The zero-point energy $E_0$ of the molecule is 
 \begin{equation}
 E_0=G \left(\frac{1}{2} \right) = \frac{1}{2}\omega_e  - \frac{1}{4}\omega_e x_e + \frac{1}{8}\omega_e y_e + \cdots\, .
 \end{equation}
    If the vibrational energy levels (in this ground electronic state) are referred to this lowest  energy level as zero, we put 
 \begin{equation}
  E_n-E_0=G\left(n+\frac{1}{2} \right) - G \left( \frac{1}{2} \right) \equiv G_0(n) \equiv \omega_0 n- \omega_0 x_0 n^2 + \omega_0 y_0 n^3 + \cdots, 
 \end{equation}
 where $\omega_0=\omega_e - \omega_e x_e + \frac{3}{4}\omega_e y_e+ \cdots$ etc. 
   What is really observed  in absorption bands  is (in cm${}^{-1}$) $\Delta E_n\equiv E_{n+1}-E_n =   \omega_0 -\omega_0x_0 - 2 \omega_0 x_0 n$ (neglecting cubic terms), and $\Delta^2 E_{n}\equiv \Delta E_{n + 1} -\Delta E_{n} = -2\omega_0 x_0 $ (which measures the anharmonicity).
 
  {\it Consequently these measurements do not give access to the quantum ground state energy $E_0$}. But the vibrational constants $\omega_e$ and $\omega_e x_e$ (or $\omega_0$ and $\omega_0 x_0$) can be determined from the observed  positions of the infrared absorption bands.

\subsubsection{Transitions involving  two electronic states}

  All possible transitions between the different vibrational levels (with and without prime below) corresponding to the two participating electronic states (the ground and the excited electronic potential curves) give rise to  the following transition frequencies $\nu$
 \begin{align}
\label{freqdiat}
\nonumber   2\pi  \nu =&E'_{n'}-E_n = 2 \pi \nu_{g-e}  + G'(n'+1/2)- G(n+1/2)  \\
\nonumber \equiv&  2 \pi \nu_{00}  + \omega'_0 n'- \omega'_0 x'_0 {n'}^2 + \omega'_0 y'_0 {n'}^3 +\cdots +\\  
&- \left\lbrack  \omega_0 n- \omega_0 x_0 {n}^2 + \omega_0 y_0 {n}^3 + \cdots \right\rbrack\, . 
\end{align}
   Here $2 \pi \nu_{g-e}$ is the difference between the respective minima of the two considered curves $E_e(R)$, and   $\nu_{00}$ stands for the frequency of the so-called  0-0 band (transition $E_0 \to E'_0$ ). In the harmonic approximation we have
  \begin{equation}
  \label{eq:nu00}
  2\pi \nu_{00}=2 \pi \nu_{g-e}+\frac{\omega'_e-\omega_e}{2}\,.
  \end{equation}
   It is important to notice that in the earliest Bohr-Sommerfeld  quantum theory, this $\nu_{00}$ reduces to just $\nu_{g-e}$. 
 
   We should be also aware that any change in the electronic level implies a change in the force constant of $U$. The equation \eqref{freqdiat}  has to be compared with the ``band system'' obtained from observations and  empirically modeled along the same scheme.
 
\begin{figure}[!ht]
\resizebox{\hsize}{!}{\includegraphics{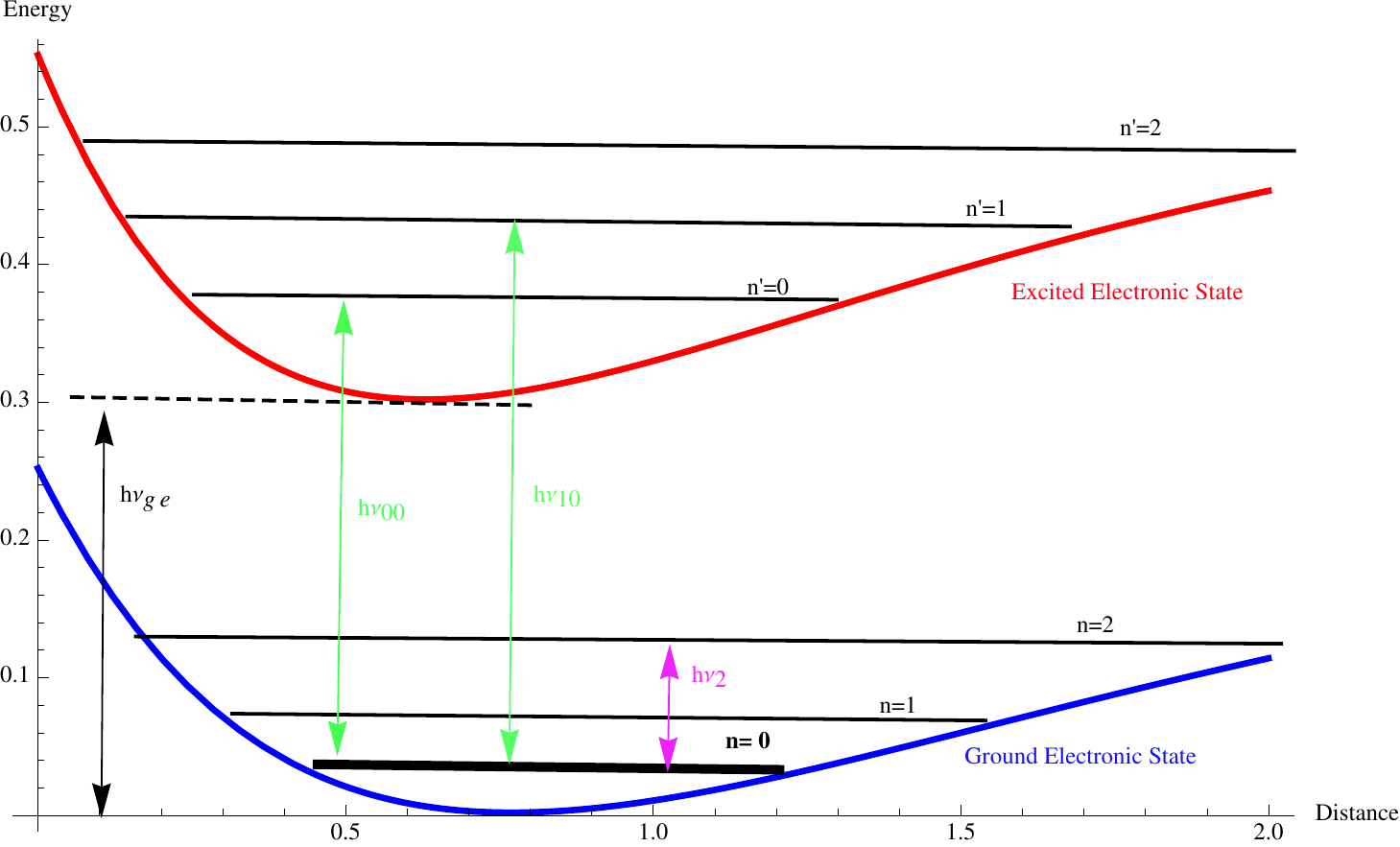}}
\caption{Sketch of the energetic diagram (arbitrary units) involving two electronic energetic curves and the corresponding vibrational levels. The thick line ($n=0$) represents the quantum ground state energy of the molecule. Some of the measured energy differences are represented: either issued from a single electronic curve ($h \nu_2$), or issued from two electronic curves ($ h \nu_{00}$ and $ h \nu_{10}$).} 
\label{figure1}
\end{figure}

     A sketch of the possible vibrational transitions is shown in Figure \ref{figure1}.

\subsection{The isotopic effect}

     When we are in presence of a gas made of isotopic molecules, like B${}^{10}$O -- B${}^{11}$O, or like HCl${}^{35}$ -- HCl${}^{37}$, assuming harmonic vibrations, we know that the (classical) vibrational pulsation $\omega_e$ is given by $\omega_e= \sqrt{\dfrac{k}{\mu}}$, where the force constant $k$ is exactly the same for different isotopic molecules, since it is determined by the electronic motion only, whereas the \textbf{reduced mass is different},
   \begin{equation}
\label{redmass}
\frac{\omega^{\mathrm{iso}}_e}{\omega_e} = \sqrt{\frac{\mu}{\mu^{\mathrm{iso}}}}\equiv \rho\,. 
\end{equation}
     Thus, in the case of a transition involving a single electronic state, the isotopic effect induces a shift $\Delta \nu $ in absorption band frequencies  given by:
   \begin{align}
\label{isoshift}
\nonumber  2 \pi \Delta \nu &= 2 \pi \nu -2 \pi \nu^{\mathrm{iso}}\\
&= (1- \rho)\left\lbrack (\omega_e -\omega_e x_e(1+\rho))n -\omega_e x_e (1+\rho)n^2\right\rbrack \approx (1- \rho) \Delta E_{n}\, , 
\end{align}
as long as $\rho$ is only slightly different from 1.
 
      In the case of transitions involving two different electronic states, and specially for $\nu_{00}$, equation \ref{eq:nu00} gives (in the harmonic approximation)
   \begin{equation}
  2 \pi \Delta \nu_{00} =(1-\rho)\frac{\omega'_e-\omega_e}{2}\, .
  \end{equation}
    Such an isotopic displacement for the 0-0 band is actually observed in a number of cases, like for  B${}^{10}$O -- B${}^{11}$O presented in Table \ref{data}. \textbf{Thus the existence of the ``zero-point'' vibration energy (the half quantum) predicted by the canonical quantization is proved}. 

\begin{table}[hbt!]
  \centering 
  \begin{tabular}{|c|c|c|c|}
\hline
  Band & Observed &  Calculated & Calculated\\
   & Isotopic   & from& from\\
   & Displacement & Quantum  & Bohr-Sommerfeld \\
   & B${}^{10}$O -- B${}^{11}$O &  Mechanics  &Theory\\
   & (cm${}^{-1}$) &  (cm${}^{-1}$)&(cm${}^{-1}$) \\
   \hline
 0-0 & -8.6 & -9.08 &0\\
  1-0 & + 26.7  & + 26.29 & + 35.69\\
 2-0& + 60.8 & + 60.36 & + 70.09\\
  3-0& + 93.6  & + 93.14 & + 103.20\\
 4-0& +125.2  & +124.63  & +135.01\\
\hline
\end{tabular}
  \caption{Isotopic displacement $2 \pi \Delta \nu_{00}$ in the $n'$-progression with $n=0$ of the $\alpha$  band of BO [According to data of Jenkins and McKellar quoted in Herzberg \cite{herzberg1989}]}\label{data}
\end{table}

\subsubsection{The predicted spectra from CS quantization}

As was shown in Section \ref{sec:genH} for the  general case, and in particular in \ref{sec:csho} for the harmonic case, the Hamiltonians provided by CS quantization differ from the ones given by canonical quantization in  additional corrections: (a) a constant energy term (proper energy) that vanishes when computing the differences between the energy levels involved in all previous formulae, (b) corrections on the potential energy that look like relativistic corrections and then are always negligible in the non-relativistic regime (corrections in $\hbar \omega /mc^2$). 

As a consequence of (a) and (b),  the final predictions on the isotopic effect provided by CS quantization are  identical to the ones given by canonical quantization, up to the negligible corrective term  $\hbar \omega /mc^2$.

\section{Conclusion}
\label{sec:concl}
The  analysis exposed in the present work shows that the predictions issued from canonical quantization and  those issued from CS quantization are perfectly compatible on a \emph{physical level}, for non-relativistic systems, even if the involved expressions differ on a mathematical level. Nevertheless this does not mean that the procedures are equivalent over the whole abstract range of mathematical observables (functions on phase space); because this set is much larger than the set of ``real" observables which are  effectively accessible to measurement.

Let us now examine two fundamental questions  raised by CS quantization. 

One can  notice that only for Hamiltonians with a
kinetic energy $p^2/2m$ the consideration of a very small  length scale leads
to a large additive constant only dependent on the mass that can be ignored. A
more general dependence (say $p^2/2m(q)$ or $\sqrt{p^2c^2 +m^2c^4}$ might lead to
quite different conclusions. 
However, our purpose is to study  fundamental Galilean  Hamiltonians and not \emph{effective} Hamiltonians (like $p^2/2m(q)$ used for example in solid state physics). Moreover these effective Hamiltonians are generally obtained through a sequence of approximations and they possess their own specific framework of validity (often they are obtained from a mixed semi-classical framework). Then there is no reason to expect a (unique) quantization procedure relevant to all these cases.
 Moreover standard CS  are adapted to the Galilean framework and not to the relativistic one. Then there is no reason to expect that  the CS quantization based on these harmonic CS be applicable to the relativistic Hamiltonian $\sqrt{p^2 c^2+m^2c^4}$. In order to work out the relativistic case we need to build CS that take into account  the Poincar\'e invariance (e.g. see \cite{aag00} and references therein). This necessary adaptation of quantization rules holds also with usual quantization, since it is well-known that the position of a particle in the relativistic context cannot be at the same time a covariant quantity and an Hermitian operator (see for instance the problems raised by  the so-called Newton-Wigner position operator \cite{hegerfeldt74,hegerfeldt80}).

The second question concerns the  applicability of CS quantization to a generic physical system. Discarding the subtle problems of relativistic corrections and other possible corrections, the canonical  quantization method produces the quantum-mechanical description for most of the systems of physical interest, e.g. for atoms, molecules.... This method seems to work so far without any limitations if only we can understand the structure of the whole system in terms of its basic constituents. Say nuclei and electrons.  Everyone agrees  on the wide range validity of the canonical quantization which should be used everywhere it is manageable and gives sound results with regard to experimental evidences.  
Nevertheless,  we know the limitations to ab initio calculations in atomic, molecular physics,  solid state physics É, imposed by  computer capabilities and theoretical complexity of $N$-body models, like the massive use of the functional density complemented with phenomenological parameters, without mentioning the inextricable tentatives for quantizing  gravity or the poor theoretical status of collective models for nuclei. So the real range of applicability  of canonical quantization is actually not so wide! 

Now, one can argue  that the CS quantization works only for a very restricted class of problems. Take, for example, the Helium atom. Can we really produce the quantum description of this system using the coherent states? And then calculate, for example, the ground state energy and the energy of the first excited state. 

Actually CS quantization represents a valid alternative for quantizing objects for which the canonical quantization does not work, like for the angle or phase  functions or distributions (in the sense of generalized functions) on the phase space. Distributions  on phase space can be used to express  geometric constraints, like for motion on manifolds, which is the case in Classical  Gravity. Furthermore, it is very motivating to build coherent states for the Kepler problem 
Our position is that there is no unique way to pass from the classical world to the quantum one, and we should be free to use all the possibilities which are offered to us and which are consistent to each other and with experimental evidences. 

\appendix

\section{Comparing Weyl and CS quantizations (and more!): a  glossary}
\label{glosspartiCSq}
\subsubsection*{Weyl-Heisenberg background}
Let $\mathcal{H}$ be a separable (complex) Hilbert space  with orthonormal basis $e_0,e_1,\dots, e_n \equiv |e_n \rangle, \dots$,  
Lowering and raising operators $a$ and $a^\dag$ are defined as
\begin{align}
   a\, \ket{e_n} & = \sqrt{n} \ket{e_{n-1}}\, , \quad  a\ket{e_0} = 0 \, \quad \mbox{(lowering operator)}\\
   a^{\dag} \, \ket{e_n} & = \sqrt{n +1} \ket{e_{n+1}} \quad \mbox{(raising operator)}\, .
\end{align}  
 To each complex number $z$ we associate the (unitary) displacement operator  or ``function $D(z)$'' :
\begin{equation}
\label{displac}
\C \ni z \mapsto D(z) = e^{z\adg -\bar z a}\, ,\quad D(-z) = (D(z))^{-1} = D(z)^{\dag}\, . 
\end{equation}
 This operator encodes the noncommutative unitary (Weyl-Heisenberg) representation of the complex plane:
\begin{equation}
\label{additiveD}
D(z)D(z') = e^{\frac{1}{2}z \circ z'}D(z+z') \,,
\end{equation}
where $z \circ z'$ is the symplectic product $z \circ z'= z\bar{z'} -\bar{z} z'= 2i\Im z\bar z'  $. 
The matrix elements of the operator $D(z)$ involve associated Laguerre polynomials $L^{(\alpha)}_n(t)$:
\begin{equation}
\label{matelD}
\lg e_m|D(z)|e_n\rg = D_{m n}(z) =  \sqrt{\dfrac{n!}{m!}}\,e^{-\vert z\vert^{2}/2}\,z^{m-n} \, L_n^{(m-n)}(\vert z\vert^{2})\, ,   \quad \mbox{for} \ m\geq n\, , 
\end{equation}
with $L_n^{(m-n)}(t) = \frac{m!}{n!} (-t)^{n-m}L_m^{(n-m)}(t)$ for $n\geq m$.
The ``parity" $\calP$ operator acts on $\mathcal{H}$ as a linear operator through
\begin{equation}
\label{parity}
\calP \ket{e_n}= (-1)^n \ket{e_n}\, , \quad \mbox{or}\ \calP= e^{i \pi a^\dag a}\,. 
\end{equation}
  This discrete symmetry verifies
\begin{align}
 \calP^2 &=\lu, \\
 \calP a \calP& =-a \;; \calP a^\dag \calP = -a^\dag,\\
  \calP D(z) \calP&=D(- z) \;.
 \end{align}

\subsubsection*{Integral formulae for $D(z)$}
\begin{itemize}

\item A first  fundamental integral: from
\begin{equation}
\label{fundintD}
\int_0^{\infty}dt\, e^{-\frac{t}{2}}\, L_n(t)  = (-2)^n\, \Rightarrow \int_{\C} \frac{d^2 z}{\pi}\, D_{m n}(z)= \delta_{mn} (-2)^m\,,
\end{equation}
it follows 
\begin{equation}
\label{fourD0}
\int_{\C} \frac{d^2 z}{\pi}\, D(z) = 2\calP\, ,
\end{equation}
\item A second fundamental integral: from (\ref{matelD}) and  the orthogonally of the associated Laguerre polynomials we obtain the ``ground state'' projector $P_0$ as the Gaussian average of $D(z)$:
\begin{equation}
\label{lapD}
 \int_{\mathbb{C}} \dfrac{d^2 z}{\pi} e^{- \frac{1}{2} |z|^2} D(z)= \ket{e_0} \bra{e_0} \,. 
\end{equation}
\item More generally for $\Re(s) < 1$
\begin{equation}
\label{lapDs}
\int_{\mathbb{C}} \dfrac{d^2 z}{\pi} e^{\frac{s}{2} |z|^2} D(z) = \dfrac{2}{1-s} \exp \left( \log \dfrac{s+1}{s-1} a^\dag a \right)\,, 
\end{equation}
where the convergence holds in norm for $\Re(s)<0$ and weakly for $0 \leq \Re(s) < 1$.
 \end{itemize}

\subsubsection*{Harmonic analysis on $\C$ and symbol calculus}
Let $f$ be a $L^1$ function on $\mathbb{C}$, its symplectic Fourier transform $\hat{f}$ is defined as:
\begin{equation}
\label{symfourier}
\hat{f}(z)=\int_{\mathbb{C}} \dfrac{d^2 \xi}{\pi} e^{ z \circ \xi } f(\xi)\, , 
\end{equation}
and  is an involution ($\hat{\hat{f}}=f$). 
Using usual symbolic integral calculus we have the Dirac-Fourier formula:
\begin{equation}
\label{diracfourier}
\int_{\mathbb{C}} \dfrac{d^2 \xi}{\pi} e^{z \circ \xi }= \pi \delta^{(2)}(z)
\end{equation}
The  resolution of the identity  follows from (\ref{fourD0}):
\begin{equation}
\label{resunweyl}
\int_{\C} \frac{d^2 z}{\pi}  \, D(z)\, 2\calP\, D(-z) = \lu\, . 
\end{equation}
This formula is at the basis of the Weyl quantization (in complex notations).
The Fourier transform of operator $D$, is easily found from (\ref{fourD0}) and Fourier transform of the addition formula (\ref{additiveD}): 
  \begin{equation}
\label{ftransfD}
\int_{\C} \frac{d^2 z'}{\pi} \,e^{(z \bar z' - \bar z z')} \, D(z') = 2 D(2z)\calP = 2 \calP D(-2z)  \,.
\end{equation}
\subsubsection*{Integral quantizations}
Let $(X,\nu)$ be a measure space. Let $\mathcal{H}$ be a (separable) Hilbert space and $X\ni x \mapsto M(x) \in \mathcal{L}(\mathcal{H})$ an $X$-labelled family of bounded operators on $\mathcal{H}$  resolving the unity $\lu$:
  \begin{equation}
\label{Psolveun}
\int_{X}\nu(dx)\, M(x) = \lu\, , 
\end{equation}
the equality being understood in a weak sense.
A formal quantization of a set of complex-valued functions $f(x)$ on $X$ is then defined by the linear map:
  \begin{equation}
\label{formquant}
f\mapsto A_f = \int_{X}\nu(dx)\, M(x) \, f(x)\, , 
\end{equation}
the definition of the operator $A_f$  being understood in a weak sense.
 In particular,  a quantization scheme for $(X,\nu)= (\C,d^2z/\pi)$ is a linear map $ f \mapsto A_f$ from the set of functions on $\mathbb{C}$ to the set of linear operators on $\mathcal{H}$. A general definition covering normal, anti-normal and Wigner-Weyl quantizations is given by
\begin{equation}
\label{quantvarpi}
A_f = \int_{\mathbb{C}} \dfrac{d^2 z}{\pi} \varpi(z) \hat{f}(-z) D(z)\, ,
\end{equation}
where $\varpi(z)$ is a weight function that specifies the type of quantization.
Equivalently we have
\begin{equation}
\label{eqquantvarpi}
A_f = \int_{\mathbb{C}} \dfrac{d^2 z}{\pi} f(z) D(z) \rho D(z)^\dag \quad \mathrm{with} \quad \rho= \int_{\mathbb{C}} \dfrac{d^2 \xi}{\pi} \varpi(\xi) D(\xi)\, . 
\end{equation}
Then
\begin{equation}
\label{quantdelta}
\rho = A_{\pi \delta^{(2)}(z)}\, .
\end{equation}
This map verifies the following important property:
\begin{equation}
\label{covquant}
A_{f(z-z_0)} = D(z_0) A_{f(z)} D(z_0)^\dag\, .
\end{equation}
Moreover we have
\begin{equation}
\label{quantvarpi}
\forall f, \, A_{f(-z)} = \calP A_{f(z)} \calP  \iff \forall z, \,  \varpi(z)=\varpi(-z)
\end{equation}
\begin{equation}
\forall f, \, A_{\overline{f(z)}} = A_{f(z)}^\dag \, \iff \, \forall z, \, \overline{\varpi(-z)}=\varpi(z)
\end{equation}
\begin{equation}
A_{1}=\lu \iff \varpi(0)=1.
\end{equation}
\subsubsection*{Regular quantizations}
We say that the quantization map $f \mapsto A_f$ is regular (in the sense it yields the canonical commutation rule $[a,a^{\dag}] = \lu$,  if the weight function $\varpi$ verifies 
\begin{equation}
\label{regvarpi}
\varpi(0)=1\, , \quad \varpi(-z)=\varpi(z)\, , \quad
\overline{\varpi(z)}=\varpi(z)\,.
\end{equation}
In that case we have 
\begin{equation}
A_{1}=\lu, \, A_{z} = a, \,  A_{\overline{f(z)}} = A_{f(z)}^\dag
\end{equation}
Therefore, for  special choices of  $\varpi$ the corresponding integral quantization is regular. Namely if we define $\varpi_s(z) = e^{s |z|^2/2}$, then $s=-1$ corresponds to the CS quantization (anti-normal), $s=0$ corresponds to the Wigner-Weyl quantization and $s=1$ is the normal quantization (which represents a limit case to be excluded in terms of integral of operators). The parameter $s$ is the same as the Cahill-Glauber parameter \cite{cahillglauber69} .
 The operator-valued measure 
\begin{equation}
\label{spovs}
\C \supset \Delta \mapsto  \int_{\Delta} \dfrac{d^2 z}{\pi} D(z) \rho_s D(z)^\dag\,, \quad \rho_s = \int_{\mathbb{C}} \dfrac{d^2 \xi}{\pi} \varpi_s(\xi) D(\xi)
\end{equation}
 is a positive operator-valued measure iff $s$ is real and $\leq -1$.  

\subsubsection*{Acknowledgements}

The authors are indebted to the Referee for valuable comments and remarks which have  contributed significantly to improve the scientific content of the manuscript.


\bibliographystyle{elsarticle-num}
\bibliography{<your-bib-database>}

\begin{thebibliography}{99}

\bibitem{mulliken25}  Mulliken R S 1925 {\it Phys. Rev.} {\bf 25} 119 and 259;  Jenkins F A, and  McKellar A 1932 {\it Phys. Rev.} {\bf42} 464;   Van Vleck  J H 1936 {\it J. Chem. Phys.} {\bf 4} 327

\bibitem{herzberg1989}  Herzberg G 1989 {\it Molecular Spectra and Molecular Structure: Spectra of Diatomic Molecules} (Krieger Pub Co; 2 edition)

\bibitem{phspq06}  Zachos C, Fairlie D, Curtright T 2006 {\it Quantum Mechanics in Phase Space: An Overview With Selected Papers}, (Singapore: World Scientific Publishing)

\bibitem{cahillglauber69} 
 Cahill K E and  Glauber R 1969   \textit{Phys. Rev.} {\bf 117} 1857

\bibitem{carnieto68} Carruthers P and Nieto M M 1968 {\it Rev. Mod. Phys.} {\bf 40}  411

\bibitem{galapon02}  Galapon  E  A  2009  in  {\it Time in quantum mechanics} Vol. 2, 25Ð63, Lecture Notes in Phys. 789, Springer, Berlin; Galapon  E  A  2002
 \textit{R. Soc. Lond. Proc. Ser. A  Math. Phys. Eng. Sci. } \textbf{458}  451

\bibitem{relat}  Berestetskii V B,  Pitaevskii L P and  Lifshitz E M 1982 {\it  Quantum Electrodynamics}
  (Butterworth-Heinemann) 2 ed.
  
 \bibitem{alienglis2005} Ali S T and Engli\v{s} M  2005  {\it Rev. Math. Phys.} {\bf 17} 391
 
 \bibitem{gazeau2009} Gazeau J P 2009 \textit{Coherent States in Quantum Physics} (Wiley-VCH)
 
 \bibitem{cotgazgor10}  Cotfas N, Gazeau J P and  G\'orska K 2010 \textit{J. Phys. A: Math. Theor.} \textbf{43}, 305304

\bibitem{bergasiyou10}  Bergeron H,  Gazeau J P, Siegl P, and A. Youssef 2010  \textit{Eur.  Phys. Lett.}   \textbf{43} 123502

 \bibitem{gasza11} Gazeau J P and  Szafraniec F H 2011
\textit{J. Phys. A: Math. Theor.}  \textbf{44} 495201

 \bibitem{bersiyou12} H. Bergeron, P. Siegl, and A. Youssef 2012 \textit{J. Phys. A: Math. Theor.} \textbf{45} 244028-1-15 (2012). 








\bibitem{klauskag1985}  Klauder J R and Skagerstam B S 1985 {\it Coherent states, applications in physics and mathematical physics} (Singapore: World scientific)

\bibitem{bechagayo2011} Bergeron H,  Chakraborty B., Gazeau J.P., and  Youssef A. 2011 \emph{Coherent state quantization of  singular observables: angle, time, and more} (in preparation)


\bibitem{lieb1973} Lieb E H 1973 {\it Commun. Math. Phys.} {\bf 31} 327

\bibitem{berezin1975} Berezin F A 1975 {\it Commun. Math. Phys.} {\bf 40} 153 

\bibitem{englis1999} Engli\v{s} M 1999   {\it  Integral Equations and Operator Theory}  {\bf 33} 426 

\bibitem{leblond1974} L\'evy-Leblond  J M 1974 {\it Riv. Nuovo Cimento (2)}  {\bf 4} 99. 

\bibitem{lande1939} Land\'{e}  A 1939 {\it Phys. Rev.} {\bf 56} 482

\bibitem{born1939} Born  M 1939 {\it Proc. R. Soc. Edinburgh} {\bf 59} 219

\bibitem{stali85} Ali S T 1985 {\it Riv. Nuovo Cim.} {\bf 8} 1

\bibitem{itzykson1980} Itzykson C and Zuber J B 1980 {\it Quantum field theory} (Singapore: McGraw-Hill).

\bibitem{aag00}  Ali S T, Antoine J-P, and  Gazeau J-P 2000  \textit{Coherent States, Wavelets and Their Generalizations},
(New York, Berlin, Heidelberg: Springer-Verlag)

\bibitem{hegerfeldt74} Hegerfeldt, G C 1974 {\it 
Phys. Rev. D} {\bf 10} 3320 

\bibitem{hegerfeldt80}Hegerfeldt, G C and Ruijsenaars N M 1980 {\it  Phys. Rev. D} {\bf 22}
377



\end{thebibliography}








\end{document}